\documentclass[floatfix,%
 aip,
 apl,
 amsmath,amssymb,
 reprint,%
final]{revtex4-1}
\usepackage{graphicx}
\usepackage{epstopdf}
\usepackage{placeins} 
\usepackage{dcolumn}
\usepackage{bm}

\usepackage[utf8]{inputenc}
\usepackage[T1]{fontenc}
\usepackage{mathptmx}
\usepackage{etoolbox}

\usepackage{graphicx}
\usepackage{amsmath,amssymb,mathrsfs,array,longtable,delarray}
\usepackage{latexsym}
\usepackage{dcolumn}
\usepackage{bm}
\usepackage{textcomp} 
\usepackage{siunitx}

\usepackage{verbatim}

\makeatletter
\def\@email#1#2{%
 \endgroup
 \patchcmd{\titleblock@produce}
  {\frontmatter@RRAPformat}
  {\frontmatter@RRAPformat{\produce@RRAP{*#1\href{mailto:#2}{#2}}}\frontmatter@RRAPformat}
  {}{}
}%
\makeatother
\usepackage{placeins}

\begin{document}

\preprint{AIP/123-QED}

\title{Hybrid Epitaxial Al/InGaAs system: Solid-state dewetting and Al facet formation}

\author{A. Elbaroudy}
\altaffiliation{corresponding author: ammaelbaroudy@uwaterloo.ca}
\affiliation{Department of Electrical and Computer Engineering, University of Waterloo, Waterloo N2L 3G1, Canada}
\affiliation{Waterloo Institute for Nanotechnology, University of Waterloo, Waterloo N2L 3G1, Canada}

\author{N. Shaw}
\affiliation{Quantum Nanofabrication and Characterization Facility (QNFCF), University of Waterloo, Waterloo ON, N2L 3G1, Canada}

\author{Sandra J. Gibson}
\affiliation{Quantum Nanofabrication and Characterization Facility (QNFCF), University of Waterloo, Waterloo ON, N2L 3G1, Canada}

\author{B. D. Moreno}
\affiliation{Canadian Light Source Inc., 44 Innovation Blvd., SK S7N 2V3, Canada}

\author{F. Sfigakis}
\affiliation{Institute for Quantum Computing, University of Waterloo, Waterloo N2L 3G1, Canada}
\affiliation{Department of Chemistry, University of Waterloo, Waterloo N2L 3G1, Canada}

\author{J. Baugh}
\affiliation{Institute for Quantum Computing, University of Waterloo, Waterloo N2L 3G1, Canada}
\affiliation{Department of Physics, University of Waterloo, Waterloo N2L 3G1, Canada}
\affiliation{Department of Chemistry, University of Waterloo, Waterloo N2L 3G1, Canada}
\affiliation{Waterloo Institute for Nanotechnology, University of Waterloo, Waterloo N2L 3G1, Canada}

\author{Z. R. Wasilewski}

\altaffiliation{zbig.wasilewski@uwaterloo.ca}
\affiliation{Institute for Quantum Computing, University of Waterloo, Waterloo N2L 3G1, Canada}
\affiliation{Department of Physics, University of Waterloo, Waterloo N2L 3G1, Canada}
\affiliation{Department of Electrical and Computer Engineering, University of Waterloo, Waterloo N2L 3G1, Canada}
\affiliation{Waterloo Institute for Nanotechnology, University of Waterloo, Waterloo N2L 3G1, Canada}

\date{\today}

\begin{abstract}

Hybrid superconductor--semiconductor platforms can host subgap electronic excitations such as Andreev bound states (ABSs); in topological regimes, a special zero-energy class, Majorana bound states (MBSs), can emerge. Here we report the growth of epitaxial Al films by molecular-beam epitaxy on $\mathrm{In_{0.75}Ga_{0.25}As}$ under near-room-temperature substrate conditions. Using a combination of AFM/SEM, cross-sectional TEM, and \emph{in situ} RHEED, we map how substrate temperature and Al deposition rate govern film morphology, continuity, and interface quality. We identify a growth window that yields continuous, superconducting Al films with an abrupt $\mathrm{Al}/\mathrm{In_{0.75}Ga_{0.25}As}$ interface and no detectable indium interdiffusion. We further investigate the thermal stability of these films under \emph{in situ} post-growth heating and \emph{ex situ} annealing following surface oxidation. For unoxidized Al, rapid surface diffusion triggers solid-state dewetting at approximately $165\,^\circ\mathrm{C}$, resulting in the formation of $\{111\}$-faceted Al islands. In contrast, the presence of a native oxide largely suppresses dewetting, with failure occurring only locally at surface defects. Annealing above the indium melting point ($156.6\,^\circ\mathrm{C}$) induces significant In surface migration in both cases, leading either to localized interfacial In inclusions beneath Al agglomerates or to uniform surface contamination at sites of localized layer breakdown. Together, these results define growth and annealing conditions for thermally robust epitaxial Al on III--V semiconductors and provide practical guidance for fabricating high-quality superconductor--semiconductor hybrid platforms for quantum devices.

\end{abstract}

\maketitle
\section{\label{sec:level1}Introduction}

Superconducting aluminum deposited \textit{in situ} on III–V semiconductors—particularly InAs and InGaAs—forms the foundation of a broad class of quantum-hardware platforms. These include Andreev spin qubits, \cite{Hays2021Coherent} gatemon and transmon devices, \cite{Casparis2018Superconducting} and related proximitized circuits, \cite{Ciaccia2023Gate} as well as topological-superconductor architectures aimed at realizing Majorana bound states. \cite{Fornieri2019Evidence, Sarma2015Majorana, Mourik2012Signatures, bergeron2024high} Implementations of Al–InAs hybrid systems typically employ either one-dimensional nanowires \cite{Mourik2012Signatures, Krogstrup2015Epitaxy} or two-dimensional electron-gas (2DEG) heterostructures. \cite{Shabani2016Two, aghaee2023inas, bergeron2024high, bergeron2023field} While nanowires enabled the first reported signatures of Majorana physics, \cite{Mourik2012Signatures} 2DEG-based platforms offer superior lithographic scalability and device density, which are essential for implementing braiding and fusion protocols. \cite{Shabani2016Two, aghaee2023inas}

Within these 2DEG architectures, material quality plays a decisive role. \textit{In situ} deposition of aluminum—made possible by compatibility with molecular beam epitaxy (MBE)—is crucial for forming atomically clean semiconductor–superconductor interfaces. \cite{Krogstrup2015Epitaxy} Such interfaces are required to achieve a hard, proximity-induced superconducting gap with strongly suppressed subgap states. \cite{chang2015hard} In addition, ultrathin Al films can sustain large in-plane critical magnetic fields, a key requirement for driving hybrid systems into the topological regime. InAs is an especially attractive semiconductor partner, as it provides the three essential ingredients for topological superconductivity: high electron mobility, strong spin–orbit coupling, and a large effective $g$-factor. \cite{aghaee2023inas, Shabani2016Two, bergeron2024high} Despite these favorable properties, the unambiguous experimental realization of a robust topological phase in Al/InAs systems remains an open challenge.

Progress toward this goal depends critically on overcoming several materials-science challenges. Although the MBE growth of InAs-based 2DEGs is now well established, supported by mature epitaxial buffer designs that yield low dislocation densities, the near-surface quantum well (QW) remains highly sensitive to disorder and oxidation. To mitigate these effects, a thin InGaAs cap layer is commonly employed to protect the InAs QW. However, this cap must remain sufficiently thin to preserve a strong superconducting proximity coupling between the Al film and the 2DEG. \cite{Shabani2016Two, cheah2023control} Alternative cap materials, such as InAlAs, have also been explored to tune Fermi-level pinning and improve electrostatic gate control. \cite{strickland2022controlling}

A second, equally important set of challenges arises during the aluminum deposition itself. Under ultra-high-vacuum conditions, Al exhibits high surface mobility on III–V surfaces, which promotes dewetting and discontinuous 3D film growth. Cooling the substrate to as low as $-40~^{\circ}\mathrm{C}$ has been reported to be an effective strategy to suppress this behavior. \cite{wang2020dependence, Shabani2016Two, cheah2023control, sarney2018reactivity, sarney2020aluminum} Beyond surface diffusion, interfacial chemical reactions can further complicate growth, leading to pitting, three-dimensional island formation, \cite{sarney2018reactivity} or indium migration into the Al layer. These effects can perturb both the superconducting properties of the film and the electrostatics of the underlying 2DEG. Although various interlayers (e.g., AlAs, GaAs, AlSb) and growth optimizations have been developed to mitigate these issues, \cite{sarney2018reactivity, telkamp2025development, sarney2020aluminum, zhang2020interface} achieving reproducible control over the final film structure remains challenging. This difficulty is exacerbated by the large lattice and crystal-structure mismatch between Al and III–V substrates, which favors the nucleation of multiple competing crystal orientations. As these domains grow and coalesce, they form polycrystalline films whose complex grain structure can degrade key superconducting properties such as the critical temperature $T_{\mathrm c}$ and the in-plane critical magnetic field. \cite{cheah2023control}

In our prior work, we demonstrated an abrupt transition from three-dimensional island growth to two-dimensional, continuous Al films on In$_{0.75}$Ga$_{0.25}$As surfaces, controlled purely by the aluminum growth rate and the substrate thermal budget. \cite{elbaroudy2024observation} By employing a high Al deposition rate of $3~\text{\AA}\,\mathrm{s}^{-1}$ at a near-room-temperature substrate temperature ($\sim14~^{\circ}\mathrm{C}$), we obtained continuous epitaxial Al films with uniformity comparable to those produced using low-temperature deposition at slow growth rate. \cite{Shabani2016Two, wang2020dependence, cheah2023control}

Here, we build on that result by further investigating the Al/In$_{0.75}$Ga$_{0.25}$As system grown at high deposition rates without an insertion layer. Using transmission electron microscopy (TEM) and electron energy-loss spectroscopy (EELS), we determine the crystal orientation of Al on InGaAs and assess indium interdiffusion into the Al film. We further characterize the superconducting properties of these layers and examine their thermal stability through post-growth annealing. Furthermore, we study the solid-state dewetting of these films upon annealing, showing how thermally driven film breakup and island faceting correlate with Indium interdiffusion. Together, these results clarify how interlayer-free, high-rate Al deposition can yield continuous superconducting films suitable for hybrid quantum devices, while delineating the thermal and kinetic limits beyond which interface reactions and dewetting degrade film quality.

\section{Experiments}

All heterostructures in this study were grown in a Veeco GEN10 MBE system on InP (100) semi-insulating Fe-doped substrates. The growth sequence for each sample consisted of a lattice-matched $\mathrm{In}_{0.53}\mathrm{Ga}_{0.47}\mathrm{As}$ buffer layer, followed by a 6~nm $\mathrm{In}_{0.75}\mathrm{Ga}_{0.25}\mathrm{As}$ layer, and ending in an epitaxial Al layer with a target thickness of 10~nm. The Al deposition parameters, specifically the growth rate and substrate rotation, were quoted for each sample as summarized in Table~\ref{tab:table1}.

This investigation builds upon our previous work, \cite{elbaroudy2024observation} in which Al layers were grown on stationary substrates. Sample G0972, used here, was grown at an Al rate of 3 $\text{\AA}\,\mathrm{s^{-1}}$, resulting in a continuous 2D Al thin film. In the present study, sample G1001 was grown at the same high rate ($3.0~\text{\AA}\,\mathrm{s^{-1}}$) but with continuous substrate rotation ($33.3~\text{rev}\,\mathrm{s^{-1}}$) to facilitate \textit{in situ} RHEED study of surface reconstruction at four different azimuths simultaneously. Finally, sample G1100—a repeat of G1001—was grown to enable a dedicated \textit{in situ} RHEED study of the Al dewetting dynamics, which was subsequently correlated with \textit{ex situ} characterization results by AFM, SEM, and TEM.

\begin{table*}
\caption{\label{tab:table1}Samples investigated in this study, summarizing key growth variables.}
\begin{ruledtabular}
\begin{tabular}{ccccc}

 Sample ID&Al thickness target (\AA)&Al growth rate (\AA/s)&Substrate rotation during Al growth\\
 \hline 

    G0972 & 100 & 3.0 & Stationary \\
    G1001 & 100 & 3.0 & 0.33 rev\,s$^{-1}$ \\
    G1100 & 100 & 3.0 & 0.33 rev\,s$^{-1}$ \\

\end{tabular}
\end{ruledtabular}
\end{table*}

For sample G1001, \textit{in situ} characterization involved RHEED monitoring of the entire growth sequence. The RHEED patterns were continuously acquired along four different substrate azimuths using a kSA~400 system. This acquisition was synchronized with the substrate's angular orientation using a kSA~RMAT optical encoder and a rotation-aligned triggering mechanism.

\begin{figure}[t]
  \centering
  \includegraphics[width=\linewidth]{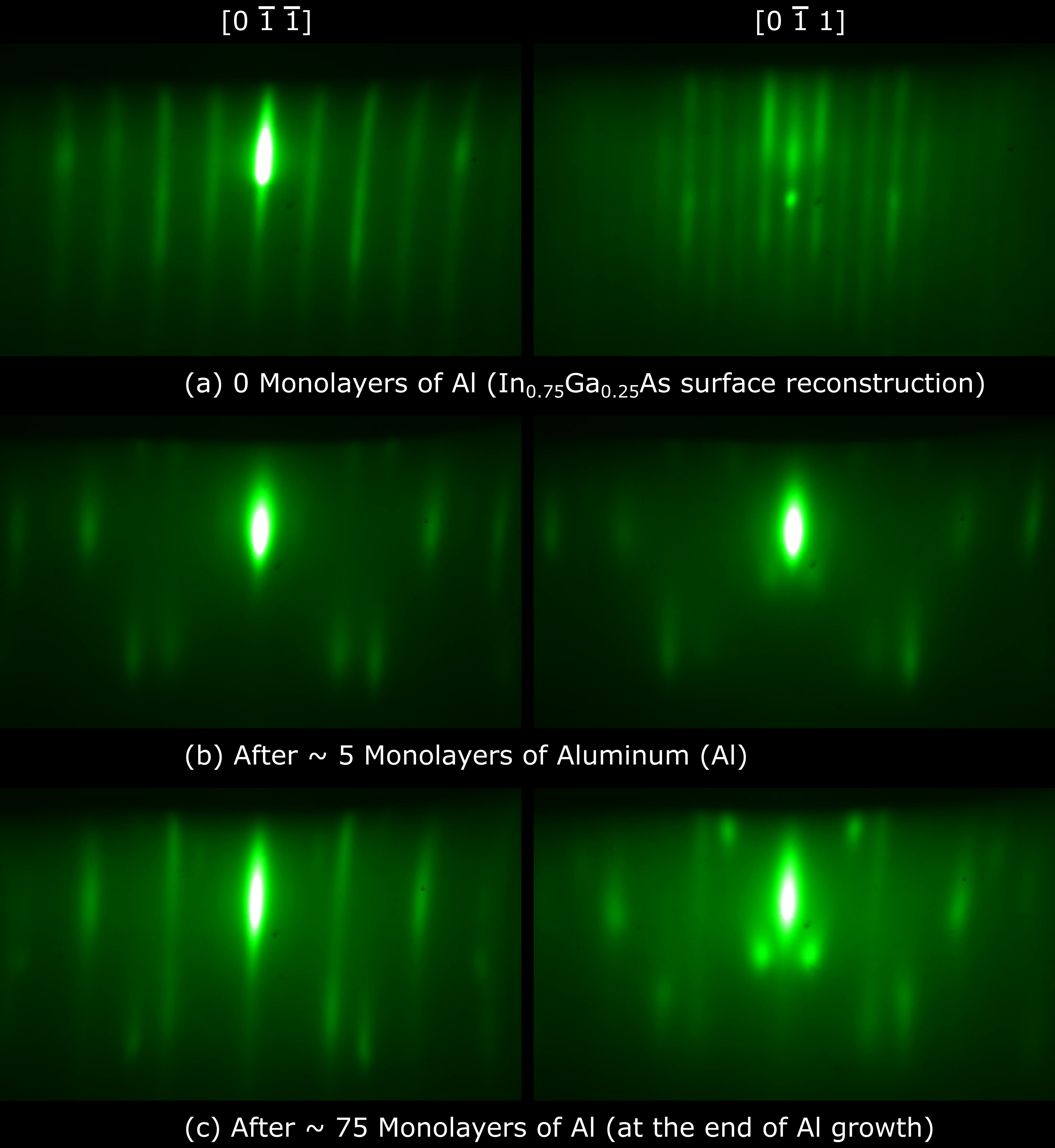}
  \caption{RHEED patterns recorded along the $[0\bar{1}\bar{1}]$ and $[0\bar{1}1]$ azimuths: (a) substrate surface after deposition of 6\,nm In$_{0.75}$Ga$_{0.25}$As; (b) after $\sim$5 monolayers of Al, showing the evolving surface morphology; and (c) upon completion of $\sim$75 monolayers of Al.}
  \label{fig:RHEEDInGaAs_Al}
\end{figure}

The substrate temperature was monitored throughout the growth process using band-edge thermometry (BET). \cite{johnson1995optical} The process began with  the oxide desorption of the $\mathrm{InP}(001)$ substrate at $535\,^{\circ}\mathrm{C}$ under $\mathrm{As}_{4}$ overpressure. Upon observation of the $(2\times4)$ reconstruction, the substrate was cooled to $470\,^{\circ}\mathrm{C}$ under $\mathrm{As}_{4}$ flux, retaining a streaky $(2\times4)$ RHEED pattern typical of the surface reconstructions of $\mathrm{InP}(001)$. \cite{labella2000reflection, dmitriev2021transformation} Once stabilized at this temperature, a $100\,\mathrm{nm}$ lattice-matched $\mathrm{In}_{0.53}\mathrm{Ga}_{0.47}\mathrm{As}$ buffer was grown under excess $\mathrm{As}_{4}$. During subsequent growth of the $\mathrm{In}_{0.75}\mathrm{Ga}_{0.25}\mathrm{As}$ layer, the RHEED pattern along the orthogonal $[0\bar{1}\bar{1}]$ and $[0\bar{1}1]$ in-plane azimuths remained streaky with a $(2\times4)$-like pattern indicative of a flat, well-ordered $(001)$ surface (Fig.~\ref{fig:RHEEDInGaAs_Al}(a)). This observation is consistent with detailed \textit{in situ} STM studies by Mirecki Millunchick et al of As-rich, strained $\mathrm{In}_{x}\mathrm{Ga}_{1-x}\mathrm{As}(001)$ surfaces.\cite{millunchick2004surface} For In-rich alloys similar to our $\mathrm{In}_{0.75}\mathrm{Ga}_{0.25}\mathrm{As}$, the surface is often not a single, uniform phase. Although the canonical As-rich structure is the $\beta_{2}(2\times4)$,  on $\mathrm{In}_{0.81}\mathrm{Ga}_{0.19}\mathrm{As}$ films revealed that a clear $(2\times4)$ RHEED pattern can correspond to a complex mosaic of $\beta_{2}(2\times4)$ domains coexisting with more disordered $(4\times3)$ and $(6\times4)$ motifs. We therefore use the ``$(2\times4)$-like'' descriptor to acknowledge that our RHEED pattern likely represents an average over these coexisting phases. This terminology also inherently includes the ambiguity between a disordered $(2\times4)$ and the ordered $\mathrm{c}(2\times8)$ reconstruction, which are RHEED-indistinguishable.

After growth of the $\mathrm{In}_{0.75}\mathrm{Ga}_{0.25}\mathrm{As}$ layer, the substrate was cooled to $400\,^{\circ}\mathrm{C}$; the As shutter and valve were then closed, and growth recipe was paused until the background pressure, dominated by As$_4$, fell below $1.2\times10^{-10}$~Torr. This was done to avoid deposition of aluminum on an As-coated surface, and suppress AlAs formation. By the following morning, the substrate temperature had passively cooled to $\sim14\,^{\circ}\mathrm{C}$. Aluminum was then deposited to a target thickness of $100~\text{\AA}$ with a growth rate of $3~ \text{\AA} ~\text{s}^{-1}$ (two cells at $1.5~ \text{\AA}~\text{s}^{-1}$ each), with the substrate rotating at $0.33 ~\text{rev}~\text{s}^{-1}$ while RHEED was recorded continuously. Upon the initiation of Al deposition, the RHEED intensity immediately decreased along both the $[0\bar{1}\bar{1}]$ and $[0\bar{1}1]$ azimuths, consistent with the formation of a partially ordered wetting layer. This was quickly followed by a broadening of the RHEED features; by $\sim$5~ML, distinct diffraction spots emerged in both azimuths, indicating island formation [Fig.~\ref{fig:RHEEDInGaAs_Al}(b)]. As growth proceeded, an abrupt transition to an Al island coalescence mode occurred. Faint two-dimensional (2D) streaks emerged, appearing first along the $[0\bar{1}\bar{1}]$ direction. These streaks progressively intensified and sharpened, eventually replacing the spotty pattern with a streaky pattern predominantly along $[0\bar{1}\bar{1}]$ by the end of the deposition [Fig.~\ref{fig:RHEEDInGaAs_Al}(c)], indicating the formation of a flat 2D Al layer.~\cite{elbaroudy2024observation}

  \begin{figure}[t]
  \centering
  \includegraphics[width=\linewidth]{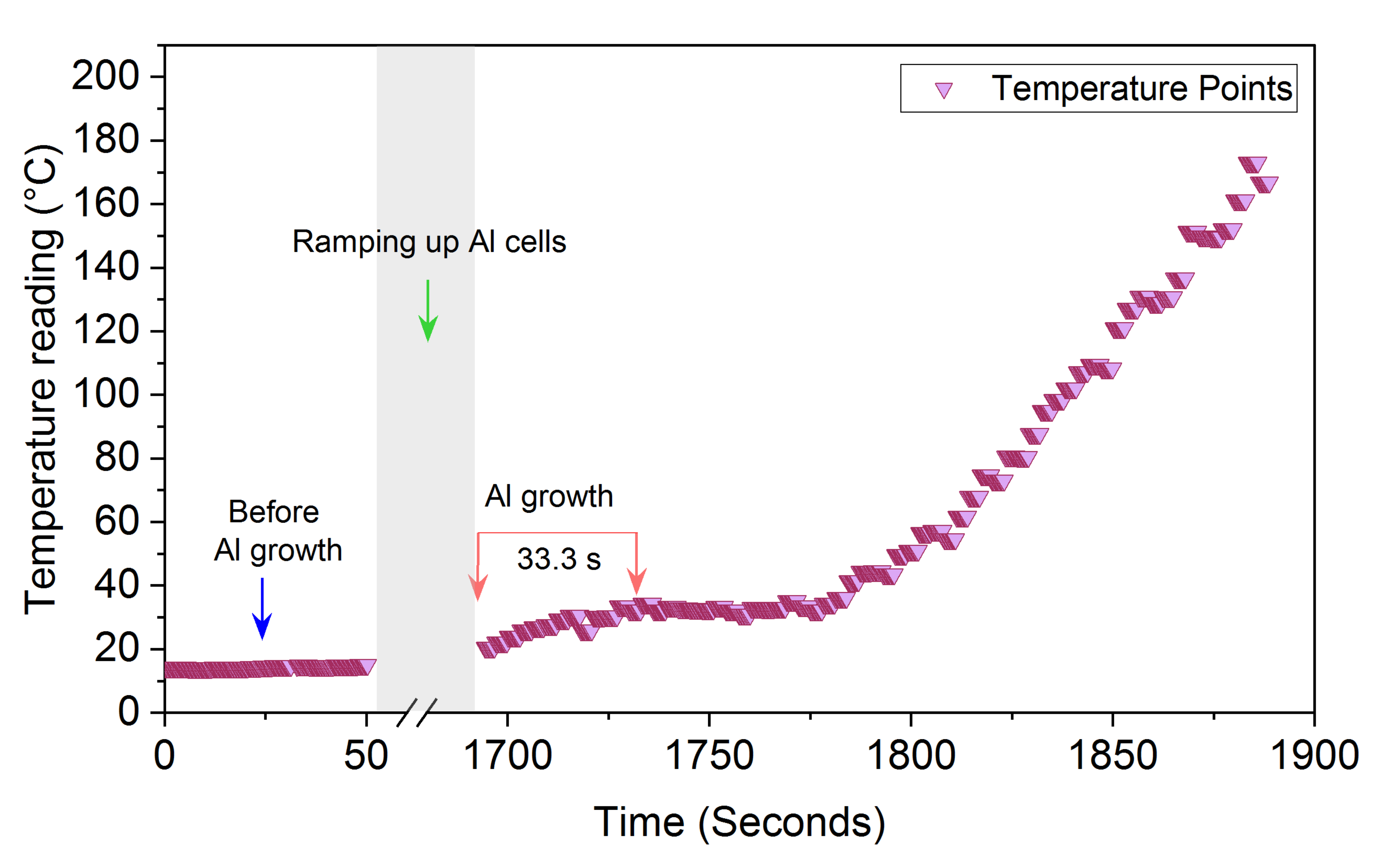}
  \caption{Substrate temperature profile for sample G1100, detailing the Al growth phase and the subsequent high-temperature dewetting ramp.}
\label{fig:subtempduringdewetting}
\end{figure}

This pronounced anisotropy is attributed to the morphology of the initial $(2\times4)$-reconstructed In$_{0.75}$Ga$_{0.25}$As surface. Surface reconstruction trenches that extend along the $[0\bar{1}\bar{1}]$ direction are preferentially filled with Al atoms, promoting the earlier establishment of a local registry and 2D ordering along this azimuth. Consequently, planar alignment along the orthogonal $[0\bar{1}1]$ direction develops more slowly. This 3D-to-2D transition sequence, observed via RHEED, is consistent with the canonical signature reported for the epitaxial growth of Al on GaAs.~\cite{tournet2016growth} Significantly, this growth-mode transition is observed only for sufficiently fast deposition rates. The 2D ordering was observed only at a high Al growth rate of $3~\text{\AA}\,\text{s}^{-1}$; at lower rates (e.g., $0.1-0.5~\text{\AA}\,\text{s}^{-1}$), the diffraction remained persistently spotty, indicative of 3D growth. This dependence of the growth rate is consistent with our previous report.~\cite{elbaroudy2024observation} The formation of a continuous uniform Al layer for the G1001 sample was further verified \textit{ex situ} with SEM.

  \begin{figure}[t]
  \centering
  \includegraphics[width=\linewidth]{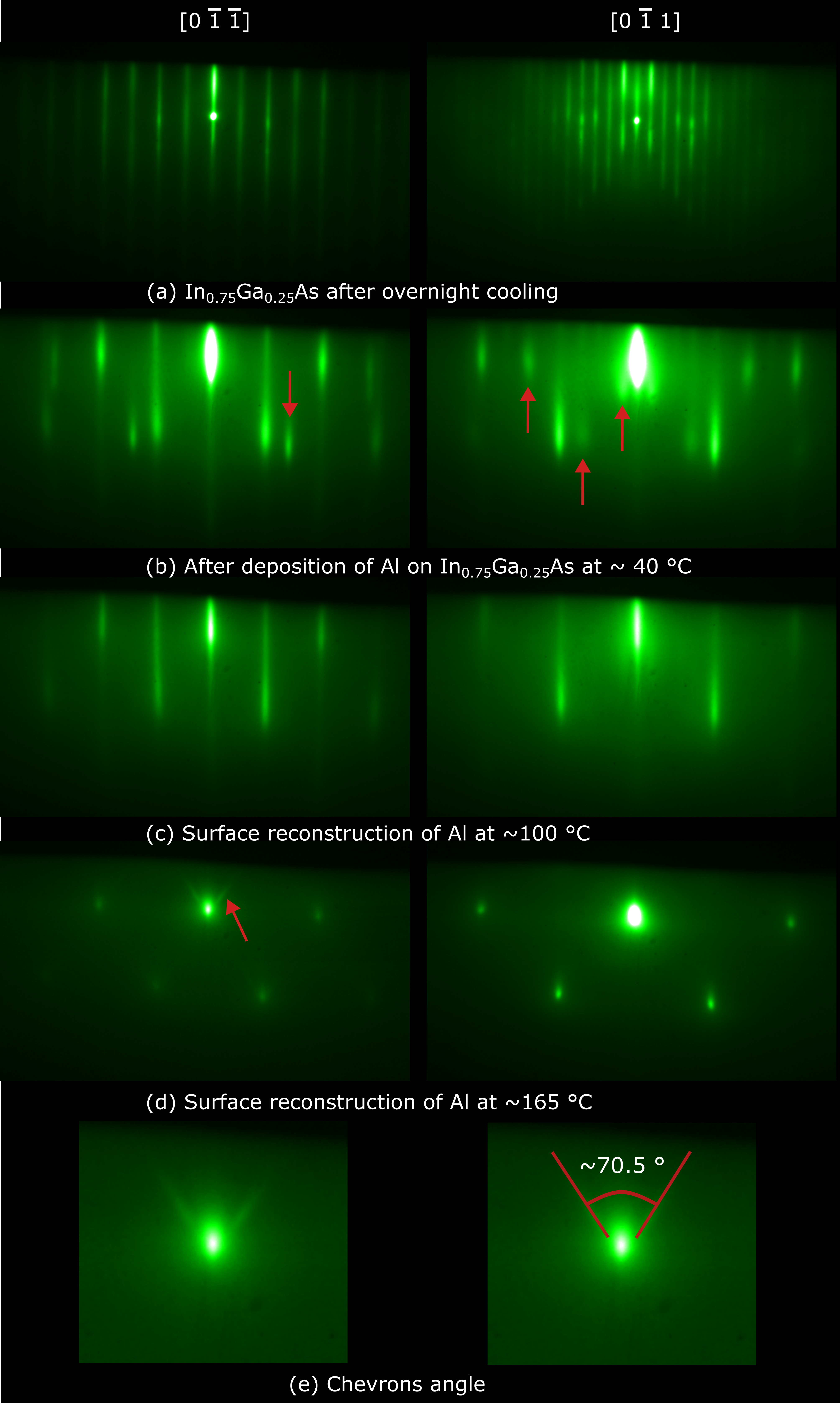}
  \caption{RHEED evolution for Al on In$_{0.75}$Ga$_{0.25}$As during post-deposition heating after $\sim$10\,nm of Al. Patterns are shown along $[0\bar{1}\bar{1}]$ (left) and $[0\bar{1}1]$ (right). 
(a) Streaky In$_{0.75}$Ga$_{0.25}$As surface after the overnight cooldown. 
(b) Immediately after Al deposition at $\sim 40\,^{\circ}\mathrm{C}$: streaky 2D Al with additional streaks from coexisting variants (red arrows). 
(c) At $\sim 100\,^{\circ}\mathrm{C}$: sharpened, purely streaky 2D pattern as a single Al orientation dominates. 
(d) At $\sim 165\,^{\circ}\mathrm{C}$: transition to spotty/3D features with chevrons, consistent with faceting during thermal roughening. (e) Chevrons angle.}
\label{fig:RHEEDAldewetting}
\end{figure}

Sample G1100 was grown using an identical procedure to G1001, with the key exception of an immediate post-deposition heating ramp, which was performed to study the solid-state dewetting of the thin Al layer on the In$_{0.75}$Ga$_{0.25}$As surface. Figure \ref{fig:subtempduringdewetting} shows the substrate temperature profile throughout the growth and annealing process for sample G1100. The process started from the baseline temperature reached overnight, followed by rapid heating of the aluminum cells to the target flux. The initial 33 seconds of the profile correspond to the deposition of $\sim$10 nm of aluminum, after which the Al cell shutters were closed, and the substrate temperature ramp was initiated. Once the temperature reached approximately $165\,^{\circ}\mathrm{C}$, the sample was cooled and removed from the system for \textit{ex situ} analysis. In the following, we describe the evolution of surface reconstruction in detail.

Much as for G1001, during the experiment, the wafer was rotated at $0.33~\mathrm{rev\,s^{-1}}$, and RHEED patterns were recorded along the $[0\bar{1}\bar{1}]$ and $[0\bar{1}1]$ azimuths. Following the overnight cooldown, the InGaAs surface initially exhibited a streaky RHEED pattern [Fig.~\ref{fig:RHEEDAldewetting}(a)]. The Al deposition was initiated at a substrate temperature of $\sim 14\,^{\circ}\mathrm{C}$, producing a streaky, 2D Al RHEED pattern [Fig.~\ref{fig:RHEEDAldewetting}(b)]. During this deposition, the temperature increased to $\sim 40\,^{\circ}\mathrm{C}$, consistent with observations during G1001 growth. Upon conclusion of the deposition, the RHEED pattern exhibited additional streaks in both azimuths, in addition to the principal integer-order streaks. These new streaks had spacings that were inconsistent with simple $1/2$ or $1/3$ fractions of the primary spacing. These features are attributed to coexisting Al epitaxial variants, such as distinct Al surface planes or in-plane rotational domains, each contributing its own integer-order rods. The coexistence of multiple Al orientations on III–V (001) surfaces, including domains with [111] and [110] normals, has been previously reported for planar Al/III–V systems.\cite{wang2020dependence} As the substrate temperature was subsequently increased during the ramp phase, the RHEED pattern evolved significantly. The extra streaks attributed to minority variants weakened and disappeared in both azimuths. By $\sim 100\,^{\circ}\mathrm{C}$, the main Al streaks sharpened into a purely streaky 2D pattern [Fig.~\ref{fig:RHEEDAldewetting}(c)]. This evolution is consistent with orientation consolidation during mild annealing, in which minority Al variants diminish as islands coalesce and recrystallize, leaving a dominant orientation. Upon further heating to $\sim165\,^{\circ}\mathrm{C}$, the 2D streaks evolved into a spotty 3D pattern, and chevrons became visible along the $[0\bar{1}\bar{1}]$ azimuth [Fig.~\ref{fig:RHEEDAldewetting}(d)]. Chevron motifs in RHEED are a known signature of faceting and 3D island formation during surface roughening. While the chevron angle reflects the evolving facet geometry, it does not by itself uniquely identify a facet family without detailed modeling. The measured chevron angle in this case is $\sim 70.5^{\circ}$ as shown in Fig.~\ref{fig:RHEEDAldewetting}(e); the specific facet indices are determined later by AFM and cross-sectional TEM analysis.

\section{Results}

\begin{figure}[t]
  \centering
  \includegraphics[width=\linewidth]{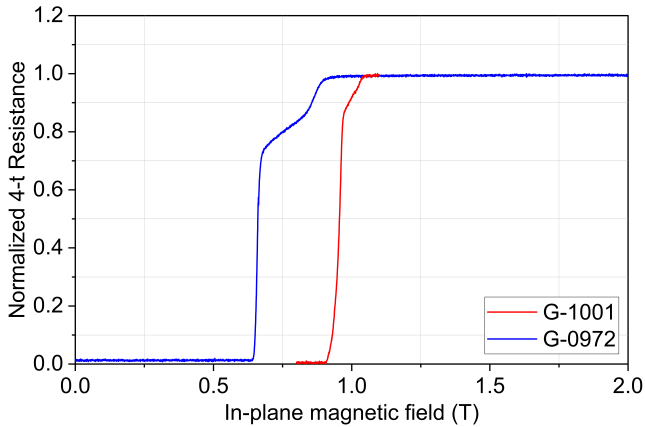}
  \caption{Critical magnetic field measurements for G1001, and G0972 samples of Al.}
  \label{fig4:SupercondG1001vsG0972}
\end{figure}

Sample G1001 features a continuous and smooth 2D Al film, with a morphology comparable to that reported in our previous work.~\cite{elbaroudy2024observation} X-ray reflectometry (XRR) data (not shown here) confirmed a film thickness of $13~\mathrm{nm}$. This deviation from the $10~\mathrm{nm}$ growth target is attributed to a higher-than-intended Al cell flux. The superconducting-to-normal transition was characterized using four-terminal AC lock-in techniques on cleaved samples ($3 \times 1~\mathrm{mm}$) in a helium-3/helium-4 dilution refrigerator (Oxford Instruments TLM, $12~\mathrm{mK}$ base temperature). This method precisely measures the film resistance, eliminating contributions from contacts and line resistances. Figure~\ref{fig4:SupercondG1001vsG0972} compares the in-plane critical magnetic field ($B_{c,\parallel}$) measurements for samples G1001 and G0972. G1001 exhibits a sharper transition at $B_{c,\parallel} = 0.97~\mathrm{T}$, a value consistent with reports for similar epitaxial Al films.~\cite{Shabani2016Two} This enhanced performance, a sharper transition, and higher $B_{c,\parallel}$ relative to G0972, is attributed to improved Al layer uniformity, which we correlate with the use of sample rotation during growth. As a type-I superconductor, an ideal Al film should show a single transition. However, both samples exhibit a second lower-field transition. We attribute this to minor thickness variations, as thicker film regions are expected to transition at a lower $B_{c,\parallel}$. This secondary transition is less pronounced in G1001, reinforcing the conclusion of better thickness uniformity. It should be mentioned that these measurements are critically sensitive to exact sample alignment in the cryostat due to the strong anisotropy between the in-plane ($B_{c,\parallel}$) and out-of-plane ($B_{c,\perp} \sim 10~\mathrm{mT}$) critical fields. Our setup lacks in-situ alignment, and even a small misalignment $\theta$ can create a perpendicular field component ($B_{\perp} = B_{\parallel} \sin\theta$) that quenches superconductivity. For example, a misalignment of just $0.5^{\circ}$ at$\ 1~\mathrm{T}$ would result in $B_{\perp} \approx 10~\mathrm{mT}$ — enough to quench the superconductivity  regardless of the $B_{c,\parallel}$ value. Consequently, while the measured $B_{c,\parallel}$ values are alignment-dependent, the relative comparison between G1001 and G0972 is valid, as both samples were mounted on the same holder.

\begin{figure}[t]
  \centering
  \includegraphics[width=\linewidth]{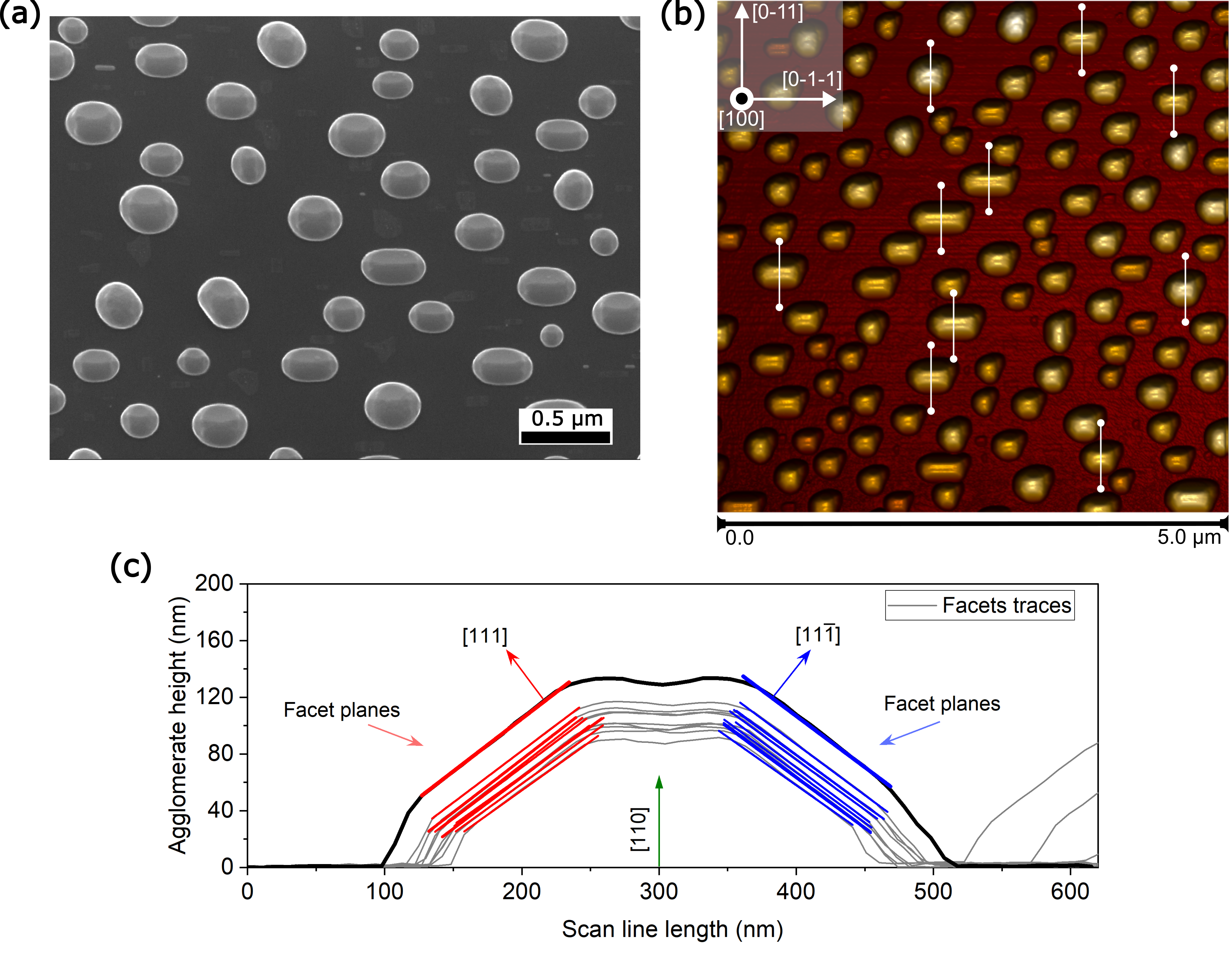}
  \caption{(a) Plan-view SEM micrograph of dewetted Al islands. (b) 3D AFM height map (5 $\mu$m $\times$ 5 $\mu$m) of the same surface. The inset indicates the in-plane [011] and [01$\bar{1}$] directions of the (100)-oriented InP substrate; white lines mark the locations of the line-scans. (c) Representative line profiles (black outline) with linear fits to the sidewalls (red/blue), indicating a consistent slope of $m \approx 0.72$, which corresponds to an inclination angle $\theta \approx 35.8^{\circ}$. The schematic overlay illustrates the proposed [110] island normal and the bounding $\{111\}$ facet planes, which are analyzed in the text.}
  \label{fig5:SEMand3DAFM}
\end{figure}

The surface dewetting initially indicated by RHEED for sample G1100 was corroborated by SEM imaging [Fig.~\ref{fig5:SEMand3DAFM}(a)], which revealed a morphology characterized by distinct, isolated Al islands. Subsequent AFM scans were used to characterize the 3D structure of these islands [Fig.~\ref{fig5:SEMand3DAFM}(b)], highlighting pronounced faceting on the island sidewalls. Height profiles extracted perpendicular to the facet ridges [Fig.~\ref{fig5:SEMand3DAFM}(c)] demonstrate a consistent sidewall inclination of $\approx 35.8^{\circ}$ relative to the substrate plane. This measured value aligns well with the theoretical angle of $35.26^{\circ}$ between the $[110]$ and $[111]$ directions characteristic of an fcc lattice, as illustrated in Fig.~\ref{fig5:SEMand3DAFM}(c). Consequently, these geometric profiles provide compelling evidence that the islands grow along the $[110]$ direction and are bounded by stable, low-energy $\{111\}$ facet planes.

\begin{figure}[t]
  \centering
  \includegraphics[width=\linewidth]{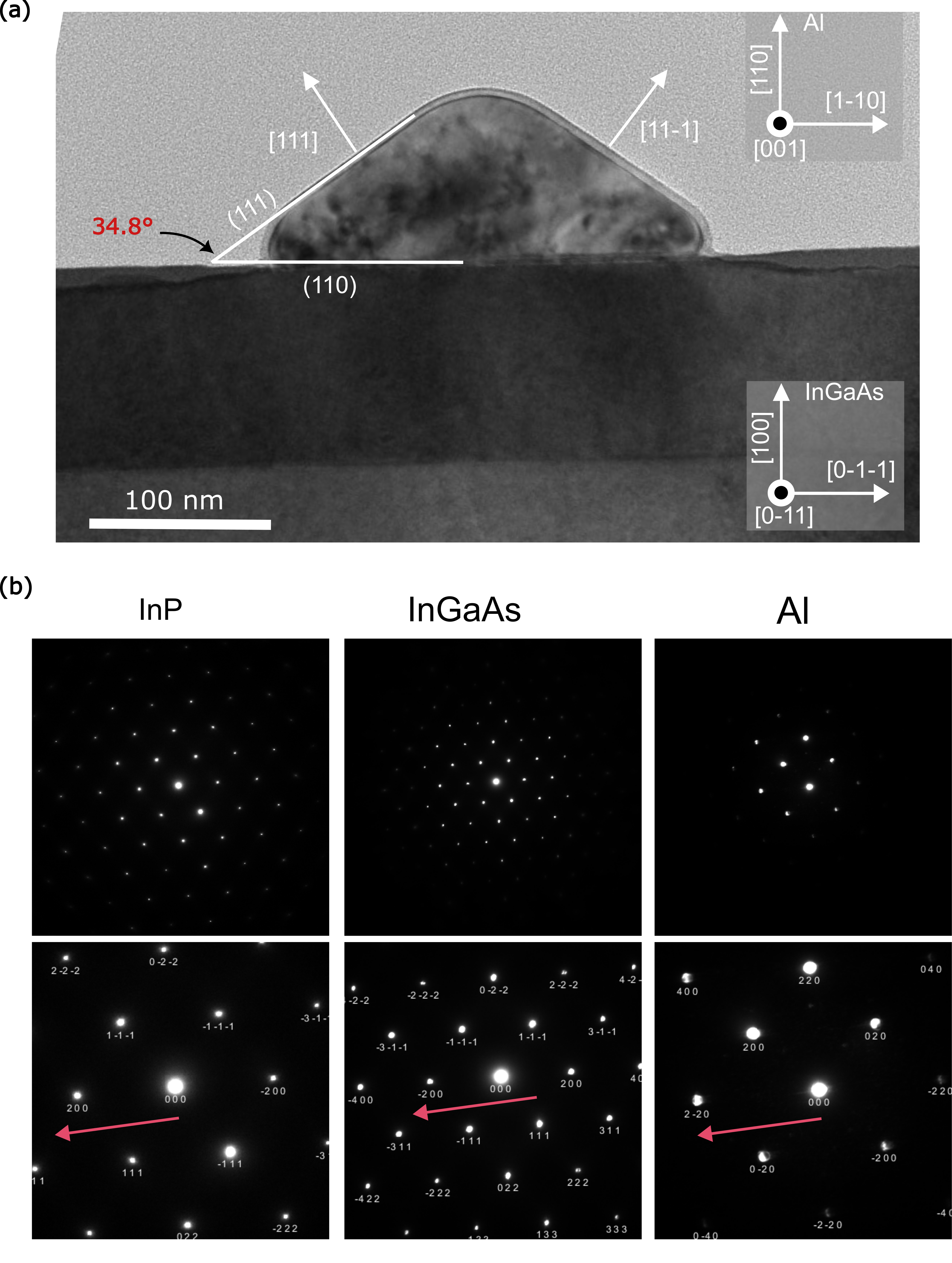}
  \caption{Representative (a) cross-sectional TEM micrograph of a faceted Al island on InGaAs from sample G1100, viewed along the Al[001] zone axis. The (110) top surface and $\{111\}$ side facets are identified. (b) NBD patterns from the InP, InGaAs, and Al layers, confirming the epitaxial relationship.}
  \label{fig6:G1100TEMandNBDs}
\end{figure}

To unambiguously index these island facets, a cross-sectional lamella from sample G1100 was prepared by Xe$^{+}$ plasma FIB for S/TEM analysis. Nano-beam diffraction (NBD) patterns acquired from the heterostructure layers [Fig.~\ref{fig6:G1100TEMandNBDs}(b)] confirm the single-crystal nature of the materials and establish the epitaxial relationship as $\text{Al}(110) \parallel \text{InGaAs}(100)$. The corresponding TEM micrograph [Fig.~\ref{fig6:G1100TEMandNBDs}(a)], viewed along the $\text{Al}[001]$ zone axis, reveals a faceted Al agglomerate. Guided by this crystallography, we identify the flat top as the (110) surface and the sidewalls as $\{111\}$ facets. This assignment is quantitatively supported by the measured sidewall inclination of $34.8^{\circ}\pm0.5^{\circ}$, which is in excellent agreement with the theoretical interplanar angle of $35.26^{\circ}$ between (110) and $\{111\}$ planes. This TEM-based identification is consistent with the AFM-measured slopes ($\theta \approx 35.8^{\circ}$) and is further corroborated by the faceting observed via RHEED. Such $\{111\}$ faceting is thermodynamically expected, as these surfaces represent the lowest-energy termination for face-centered cubic Al.

\begin{figure}[t]
  \centering
  \includegraphics[width=\linewidth]{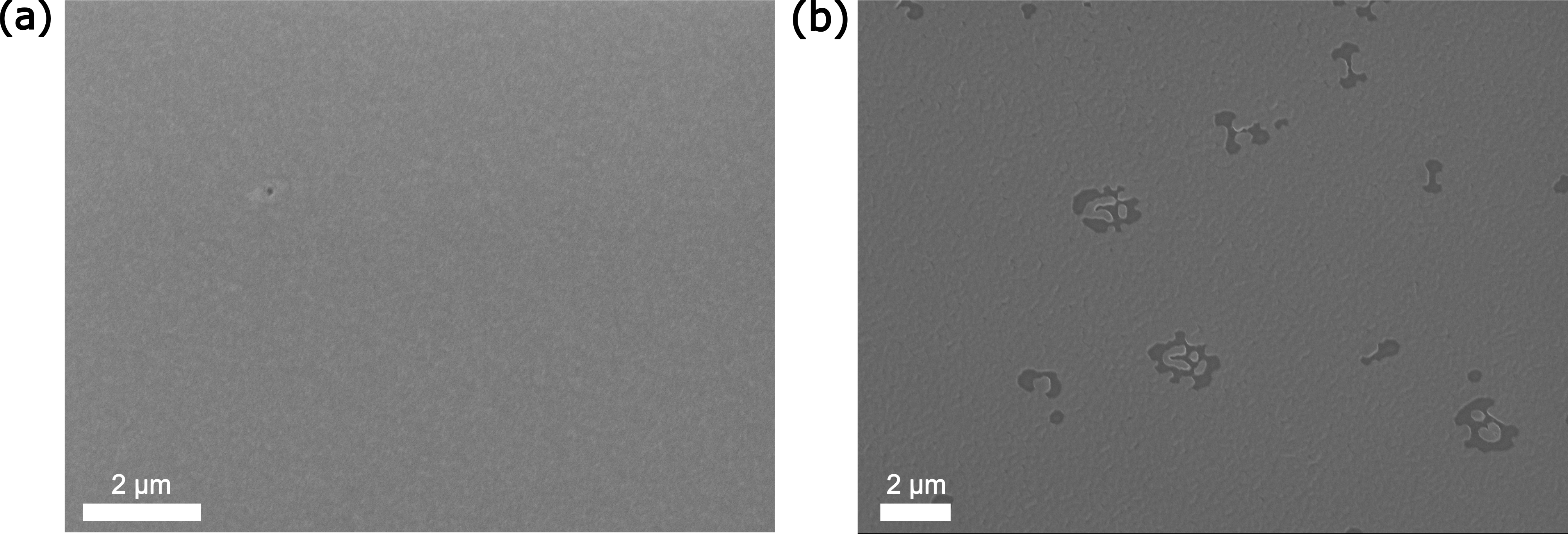}
  \caption{SEM micrographs of sample G1001. (a) As grown (no post-growth anneal). (b) After 1\,h anneal at \(200^{\circ}\mathrm{C}\) under high vacuum (\(\sim 3\times10^{-8}\) Torr). Partial dewetting is evident only after the \(200^{\circ}\mathrm{C}\) anneal.}
  \label{fig7:SEMG1001_0and200}
\end{figure}

Next, we investigated the thermal stability of the Al layer in sample G1001. This sample, which featured a continuous Al layer, was exposed to the atmosphere, allowing a native AlO$_x$ passivation layer to form. Two small pieces from sample G1001 were then annealed separately in the MBE load-lock (base pressure $\sim 3\times10^{-8}$~Torr) for one hour at nominal setpoints of $100^{\circ}\mathrm{C}$ and $200^{\circ}\mathrm{C}$, respectively. Since the radiative heater thermocouple does not make contact with the wafer, the true wafer temperature ($T$) was estimated by placing small pieces of high-purity In ($T_{\mathrm{melt}} = 156.6^{\circ}\mathrm{C}$) and Sn ($T_{\mathrm{melt}} = 231.9^{\circ}\mathrm{C}$) near the samples on the sapphire diffuser plate. The $100^{\circ}\mathrm{C}$ anneal did not melt the In, while the $200^{\circ}\mathrm{C}$ anneal melted the In but not the Sn, thus constraining the higher-temperature process window to $156.6^{\circ}\mathrm{C} < T < 231.9^{\circ}\mathrm{C}$. The resulting surface morphologies are shown in Fig.~\ref{fig7:SEMG1001_0and200}. The as-grown G1001 film [Fig.~\ref{fig7:SEMG1001_0and200}(a)] appears continuous and relatively featureless. The $100^{\circ}\mathrm{C}$ anneal ($T < 156.6^{\circ}\mathrm{C}$) produced no observable change under SEM (not shown). In contrast, the higher-temperature anneal ($156.6^{\circ}\mathrm{C} < T < 231.9^{\circ}\mathrm{C}$) induced partial solid-state dewetting at several spots on the wafer, as seen in Fig.~\ref{fig7:SEMG1001_0and200}(b).

This annealing was performed on a film capped with a native AlO$_x$ layer. This surface oxide acts to suppress Al surface diffusion, kinetically hindering the long-range material transport required for the organized facet formation observed in sample G1100. We therefore attribute the observed \textit{ex situ} dewetting to strain relaxation of the Al on the InGaAs layer. This process likely initiates not via surface transport, but at pre-existing structural defects within the Al film—such as pinholes or point defects originating from the InGaAs cap. This stands in distinct contrast to sample G1100, where the high mobility of surface Al adatoms enabled the system to reach a low-energy state by forming well-defined, faceted crystallites.

\begin{figure}[t]
  \centering
  \includegraphics[width=\linewidth]{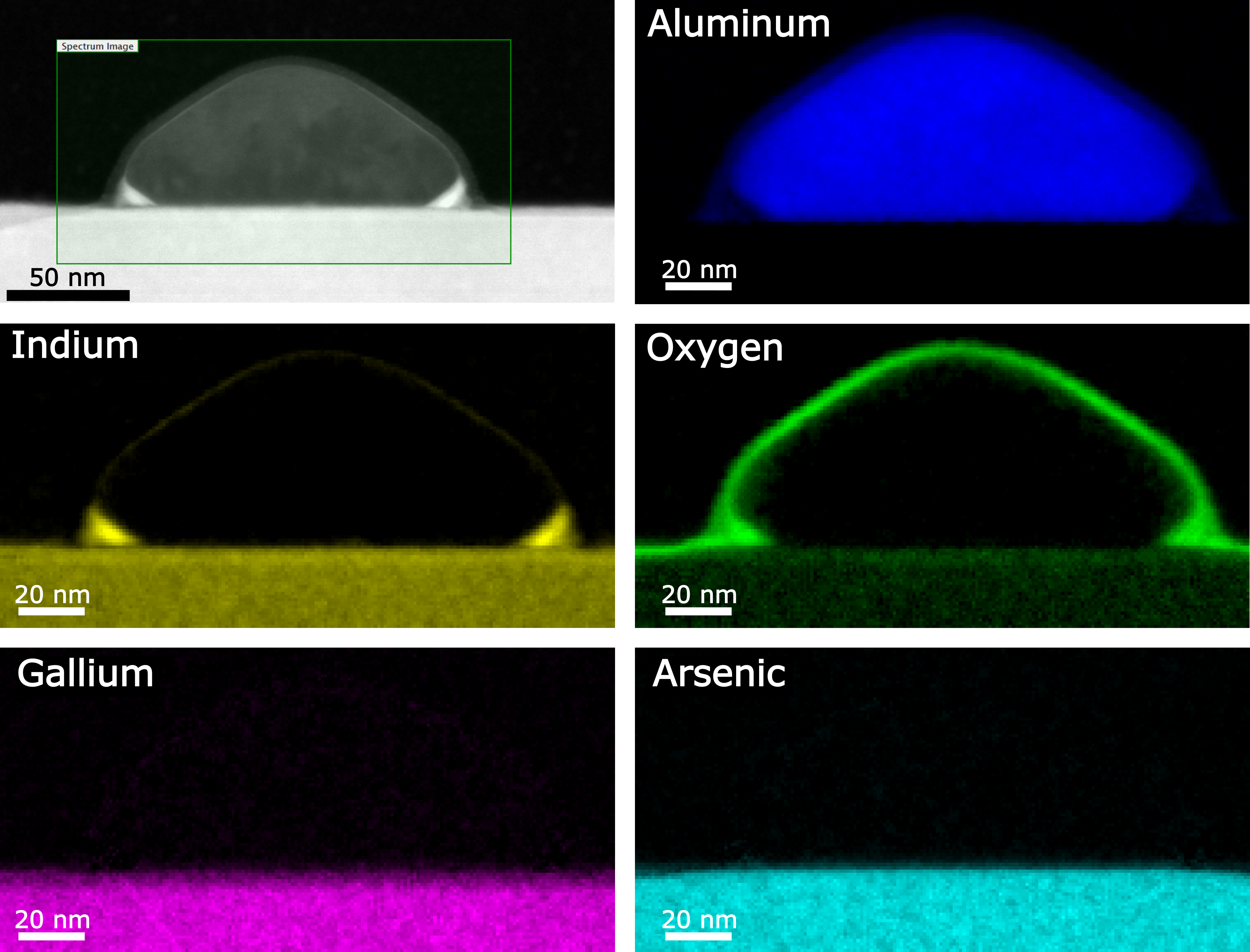}
  \caption{STEM-EELS analysis of a dewetted Al island on InGaAs from sample G1100. The elemental maps confirm the Al island (blue), a thin encapsulating oxygen shell (green), and the InGaAs substrate (Ga in pink, As in cyan). Sharp, triangular Indium inclusions (yellow) are highly localized at the interface corners, indicating a localized diffusion mechanism.}
  \label{fig8:EELS_G1100}
\end{figure}

\begin{figure}[t]
  \centering
  \includegraphics[width=\linewidth]{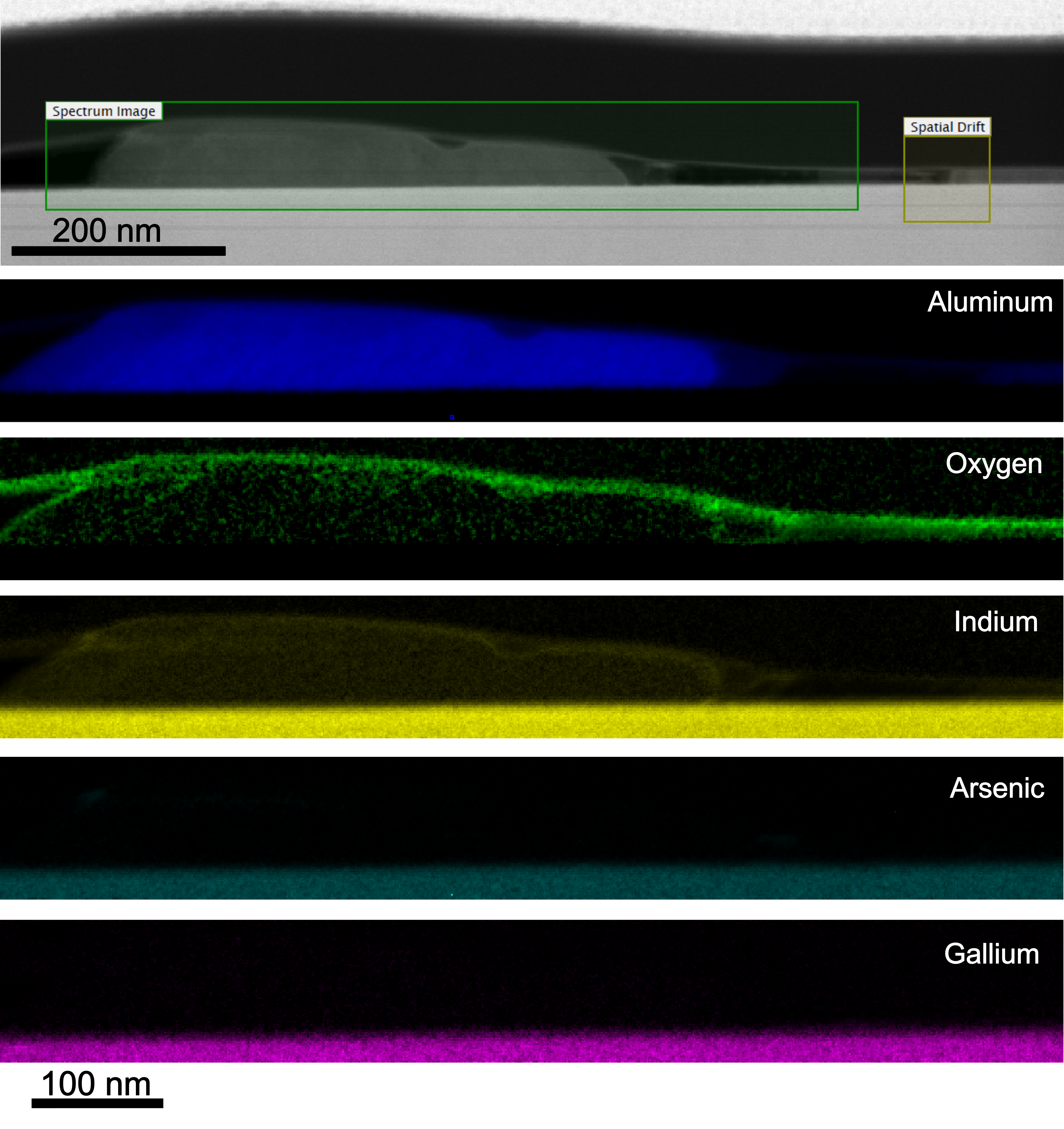}
  \caption{STEM-EELS analysis of sample G1001, annealed at 200 °C. The elemental maps for Al (blue), O (green), In (yellow), Ga (pink), and As (cyan) reveal a different diffusion profile. The Indium signal is not localized, but is broadly distributed and strongly correlated with a thick oxygen layer.}
  \label{fig9:EELS_G1001afteranneal}
\end{figure}

A distinctive characteristic observed in both the faceted G1100 sample and the dewetted G1001 sample is the presence of indium interdiffusion at the Al/InGaAs interface. Electron Energy Loss Spectroscopy (EELS) mapping of a cross-sectional lamella from sample G1100 (Fig.~\ref{fig8:EELS_G1100}) reveals well-defined, wedge-shaped indium inclusions at the base of the Al islands, directly at the substrate interface. Although these triangular inclusions were not visible on every island in the scan, this may be an artifact of the specific cross-sectional plane captured by the lamella rather than an absence of the feature in some Al agglomerates. The elemental maps further revealed that oxygen forms a thin, distinct shell encapsulating the Al islands, whereas neither gallium nor arsenic was detected within the Al layer itself. In contrast, the cross-section of sample G1001 (annealed at $200^{\circ}\mathrm{C}$) exhibits a markedly different diffusion profile (Fig.~\ref{fig9:EELS_G1001afteranneal}). In this sample, the indium signal is not confined to discrete interfacial inclusions; instead, it is distributed uniformly across the Al layer, coinciding with a significantly thicker and more dominant oxygen layer.

\section{Discussion}

We have demonstrated that high-growth-rate (3 \AA/s), near-room-temperature (14$^{\circ}$C) MBE can produce smooth, continuous epitaxial Al films on InGaAs (G1001). These films exhibit an abrupt interface and a critical magnetic field of $B_{c,||} \geq 0.97$~T. Consequently, this method presents a viable and accessible alternative to complex low-temperature growth protocols of Al on semiconductors.

The central finding of this work is the observation of solid-state dewetting of Al on InGaAs at approximately 165$^{\circ}$C, significantly below the bulk melting point of Aluminum. We attribute the dewetting of the unoxidized film (G1100) to rapid surface mass transport, which efficiently minimizes surface energy through the formation of discrete, $\{111\}$-faceted islands. In contrast, the \textit{ex situ} annealed film (G1001), capped with a native AlO$_x$ layer, exhibits suppressed dewetting kinetics. The oxide layer acts as a barrier to surface diffusion, forcing a distinct failure mode driven by void nucleation at defects and film retraction rather than organized faceting. Thus, while oxide passivation improves thermal robustness, it does not eliminate the thermodynamic driving force for dewetting at what may be surface-defect locations. This behavior is consistent with the modern understanding of solid-state dewetting as a mass transport process driven by the reduction of surface/interface free energies, which can occur well below bulk melting temperature.~\cite{leroy2016control,wang2015sharp,zucker2013model}

As revealed by EELS microscopy, such a low annealing temperature was sufficient to drive localized Indium interdiffusion in the \textit{in situ} G1100 sample. We propose that during the dewetting process, the migration and high strain of the Al layer facilitate the gathering of Indium from the InGaAs substrate, causing it to recrystallize as localized inclusions at the facet edges. Our observation of triangular Indium inclusions in sample G1100 is consistent with atomic-resolution imaging recently reported by Telkamp et al..~\cite{telkamp2025development} They used BF/HAADF-STEM and EELS to identify the tendency for localized indium diffusion into the aluminum layer from the underlying InGaAs capping. Their work indicates that this interdiffusion stems from the direct interaction between pure Al and the InGaAs, as they observed these features even without post-growth annealing. Furthermore, Telkamp et al. found that a thin GaAs capping layer of optimal thickness ($\approx 2$~ML) effectively blocks In diffusion, preserving interface quality and superconducting properties. Collectively, these findings highlight that thin Al layers on InGaAs are susceptible to dewetting and interdiffusion even when grown as continuous 2D films, suggesting a fundamental thermal instability well below the Al melting point.

\section{Summary and Conclusion}

In summary, we have demonstrated that high-growth-rate ($3~\text{\AA}\text{s}^{-1}$), near-room-temperature ($14^{\circ}\text{C}$) MBE yields continuous superconducting epitaxial Al films on InGaAs(001) with abrupt interfaces and high critical magnetic field along the layer surface. This approach offers an accessible alternative to complex low-temperature growth protocols in most MBE systems. Moreover, despite the high structural quality of the as-grown films, we identified an instability of such thin Al layers at temperatures well below the bulk melting point of aluminum. 

Our analysis identifies two distinct dewetting modes governed by thermal processing and surface condition. In unoxidized films, rapid surface diffusion initiates solid-state dewetting at approximately $165^{\circ}\text{C}$, leading to the formation of $\{111\}$-faceted Al islands. In contrast, the presence of a native oxide largely stabilizes the Al layer, suppressing dewetting over most of the surface. Localized dewetting is nevertheless observed, likely originating at pre-existing surface defects.

Annealing above the indium melting point ($156.6^{\circ}\text{C}$) induces pronounced In surface migration in both cases. This results in localized interfacial In inclusions at the base of Al agglomerates for UHV annealing performed immediately after continuous Al deposition, whereas in oxidized Al films it leads to uniform surface contamination by In at sites of localized layer failure.

\section{Acknowledgments}

We acknowledge support from funding sources and institutions. This research was made possible partly by the financial aid
from the Canada First Research Excellence Fund, specifically the
Transformative Quantum Technologies (TQT) program, and the
Natural Sciences and Engineering Research Council (NSERC) of
Canada. The research was primarily conducted in the QNC-MBE
lab. Additionally, significant use was made of the University of
Waterloo’s QNFCF Facility. The realization of this infrastructure owes much to the substantial contributions from CFREF-TQT, the Canada Foundation for Innovation (CFI), Innovation, Science and Economic Development Canada (ISED), the Ontario Ministry of Research and Innovation, as well as the generous support of Mike and Ophelia Lazaridis. The assistance and resources provided by all these entities are deeply appreciated and have been instrumental in the progression of this work.

\section{AUTHOR DECLARATIONS}
\textbf{Conflict of Interest}

The authors have no conflicts to disclose

\section{DATA AVAILABILITY}
The data that support the findings of this study are available
from the corresponding author upon reasonable request.

\section{References}

\nocite{*}
\bibliography{aipsamp}

@article{Hays2021Coherent,
  author = {Hays, M. and Fatemi, V. and Bouman, D. and Cerrillo, J. and Diamond, S. and Serniak, K. and Connolly, T. and Krogstrup, P. and Nygård, J. and Yeyati, A. Levy and Geresdi, A. and Devoret, M. H.},
  title = {Coherent manipulation of an Andreev spin qubit},
  journal = {Science},
  volume = {373},
  issue = {6553},
  pages = {430--433},
  year = {2021},
  publisher = {American Association for the Advancement of Science (AAAS)},
  doi = {10.1126/science.abf0345},
}

@article{Casparis2018Superconducting,
  author = {Casparis, L. and Connolly, M. R. and Kjaergaard, M. and Pearson, N. J. and Kringhøj, A. and Larsen, T. W. and Kuemmeth, F. and Wang, T. and Thomas, C. and Gronin, S. and Gardner, G. C. and Manfra, M. J. and Marcus, C. M. and Petersson, K. D.},
  title = {Superconducting gatemon qubit based on a proximitized two-dimensional electron gas},
  journal = {Nature Nanotechnology},
  volume = {13},
  issue = {10},
  pages = {915--919},
  year = {2018},
  publisher = {Nature Publishing Group},
  doi = {10.1038/s41565-018-0207-y},
}

@article{Ciaccia2023Gate,
  author = {Ciaccia, Carlo and Haller, Roy and Drachmann, Asbjørn C. C. and Lindemann, Tyler and Manfra, Michael J. and Schrade, Constantin and Schönenberger, Christian},
  title = {Gate-tunable Josephson diode in proximitized InAs supercurrent interferometers},
  journal = {Physical Review Research},
  volume = {5},
  issue = {3},
  pages = {033131},
  year = {2023},
  publisher = {American Physical Society (APS)},
  doi = {10.1103/PhysRevResearch.5.033131},
}

@article{Mourik2012Signatures,
  author = {Mourik, V. and Zuo, K. and Frolov, S. M. and Plissard, S. R. and Bakkers, E. P. A. M. and Kouwenhoven, L. P.},
  title = {Signatures of Majorana fermions in hybrid superconductor-semiconductor nanowire devices},
  journal = {Science},
  volume = {336},
  issue = {6084},
  pages = {1003--1007},
  year = {2012},
  publisher = {American Association for the Advancement of Science (AAAS)},
  doi = {10.1126/science.1222362},
}

@article{Sarma2015Majorana,
  author = {Sarma, S. Das and Freedman, M. and Nayak, C.},
  title = {Majorana zero modes and topological quantum computation},
  journal = {npj Quantum Information},
  volume = {1},
  issue = {1},
  pages = {1--13},
  year = {2015},
  publisher = {Nature Publishing Group},
  doi = {10.1038/npjqi.2015.1},
}

@article{Fornieri2019Evidence,
  author = {Fornieri, A. and Whiticar, A. M. and Setiawan, F. and Portolés, E. and Drachmann, A. C. C. and Keselman, A. and Gronin, S. and Thomas, C. and Wang, T. and Kallaher, R. and Gardner, G. C. and Manfra, M. J. and Krogstrup, P. and Nichele, F. and Marcus, C. M.},
  title = {Evidence of topological superconductivity in planar Josephson junctions},
  journal = {Nature},
  volume = {569},
  issue = {7754},
  pages = {89--92},
  year = {2019},
  publisher = {Nature Publishing Group},
  doi = {10.1038/s41586-019-1068-8},
}

@article{Krogstrup2015Epitaxy,
  author = {Krogstrup, P. and Ziino, N. L. B. and Chang, W. and Albrecht, S. M. and Madsen, M. H. and Johnson, E. and Nygård, J. and Marcus, C. M. and Jespersen, T. S.},
  title = {Epitaxy of semiconductor–superconductor nanowires},
  journal = {Nature Materials},
  volume = {14},
  issue = {4},
  pages = {400--406},
  year = {2015},
  publisher = {Nature Publishing Group},
  doi = {10.1038/nmat4173},
}

@article{Shabani2016Two,
  author = {Shabani, J. and Kjærgaard, M. and Suominen, H. J. and Kim, Younghyun and Nichele, F. and Pakrouski, K. and Stankevic, T. and Lutchyn, R. M. and Krogstrup, P. and Feidenhans'l, R. and Kraemer, S. and Nayak, C. and Troyer, M. and Marcus, C. M. and Palmstrøm, C. J.},
  title = {Two-dimensional epitaxial superconductor-semiconductor heterostructures: A platform for topological superconducting networks},
  journal = {Physical Review B},
  volume = {93},
  issue = {15},
  pages = {155402},
  year = {2016},
  publisher = {American Physical Society (APS)},
  doi = {10.1103/PhysRevB.93.155402},
}

@article{aghaee2023inas,
  title={InAs-Al hybrid devices passing the topological gap protocol},
  author={Aghaee, Morteza and Akkala, Arun and Alam, Zulfi and Ali, Rizwan and Alcaraz Ramirez, Alejandro and Andrzejczuk, Mariusz and Antipov, Andrey E and Aseev, Pavel and Astafev, Mikhail and Bauer, Bela and others},
  journal={Physical Review B},
  volume={107},
  number={24},
  pages={245423},
  year={2023},
  publisher={APS}
}

@article{chang2015hard,
  title={Hard gap in epitaxial semiconductor--superconductor nanowires},
  author={Chang, W and Albrecht, SM and Jespersen, TS and Kuemmeth, Ferdinand and Krogstrup, P and Nyg{\aa}rd, J and Marcus, Charles M},
  journal={Nature nanotechnology},
  volume={10},
  number={3},
  pages={232--236},
  year={2015},
  publisher={Nature Publishing Group UK London}
}

@article{strickland2022controlling,
  title={Controlling Fermi level pinning in near-surface InAs quantum wells},
  author={Strickland, William M and Hatefipour, Mehdi and Langone, Dylan and Farzaneh, SM and Shabani, Javad},
  journal={Applied Physics Letters},
  volume={121},
  number={9},
  year={2022},
  publisher={AIP Publishing}
}

@article{cheah2023control,
  title={Control over epitaxy and the role of the InAs/Al interface in hybrid two-dimensional electron gas systems},
  author={Cheah, Erik and Haxell, Daniel Z and Schott, R{\"u}diger and Zeng, Peng and Paysen, Ekaterina and ten Kate, Sofieke C and Coraiola, Marco and Landstetter, Max and Zadeh, Ali B and Trampert, Achim and others},
  journal={Physical Review Materials},
  volume={7},
  number={7},
  pages={073403},
  year={2023},
  publisher={APS}
}

@article{wang2020dependence,
  title={The dependence of aluminum lattice orientation on semiconductor lattice parameter in planar InAs/Al hybrid heterostructures},
  author={Wang, Tiantian and Thomas, Candice and Diaz, Rosa E and Gronin, Sergei and Passarello, Donata and Gardner, Geoffrey C and Capano, Michael A and Manfra, Michael J},
  journal={Journal of Crystal Growth},
  volume={535},
  pages={125570},
  year={2020},
  publisher={Elsevier}
}

@article{telkamp2025development,
  title={Development of a Nb-Based Semiconductor-Superconductor Hybrid 2DEG Platform},
  author={Telkamp, Sjoerd and Antonelli, Tommaso and Todt, Clemens and Hinderling, Manuel and Coraiola, Marco and Haxell, Daniel and Kate, Sofieke C ten and Sabonis, Deividas and Zeng, Peng and Schott, R{\"u}diger and others},
  journal={Advanced Electronic Materials},
  volume={11},
  number={7},
  pages={2400687},
  year={2025},
  publisher={Wiley Online Library}
}

@article{bergeron2024high,
  title={High transparency induced superconductivity in field effect two-dimensional electron gases in undoped InAs/AlGaSb surface quantum wells},
  author={Bergeron, E Annelise and Sfigakis, F and Elbaroudy, A and Jordan, AWM and Thompson, F and Nichols, George and Shi, Y and Tam, Man Chun and Wasilewski, ZR and Baugh, J},
  journal={Applied Physics Letters},
  volume={124},
  number={22},
  year={2024},
  publisher={AIP Publishing}
}

@article{sarney2018reactivity,
  title={Reactivity studies and structural properties of Al on compound semiconductor surfaces},
  author={Sarney, Wendy L and Svensson, Stefan P and Wickramasinghe, Kaushini S and Yuan, Joseph and Shabani, Javad},
  journal={Journal of Vacuum Science \& Technology B},
  volume={36},
  number={6},
  year={2018},
  publisher={AIP Publishing}
}

@article{sarney2020aluminum,
  title={Aluminum metallization of III--V semiconductors for the study of proximity superconductivity},
  author={Sarney, Wendy L and Svensson, Stefan P and Leff, Asher C and Schiela, William F and Yuan, Joseph O and Dartiailh, Matthieu C and Mayer, William and Wickramasinghe, Kaushini S and Shabani, Javad},
  journal={Journal of Vacuum Science \& Technology B},
  volume={38},
  number={3},
  year={2020},
  publisher={AIP Publishing}
}

@article{elbaroudy2024observation,
  title={Observation of an abrupt 3D-2D morphological transition in thin Al layers grown by MBE on InGaAs surface},
  author={Elbaroudy, A and Khromets, B and Sfigakis, F and Bergeron, E and Shi, Y and Tam, MCA and Nichols, G and Blaikie, T and Baugh, J and Wasilewski, ZR},
  journal={Journal of Vacuum Science \& Technology A},
  volume={42},
  number={3},
  year={2024},
  publisher={AIP Publishing}
}

@article{bergeron2023field,
  title={Field effect two-dimensional electron gases in modulation-doped InSb surface quantum wells},
  author={Bergeron, E Annelise and Sfigakis, F and Shi, Y and Nichols, George and Klipstein, PC and Elbaroudy, A and Walker, Sean M and Wasilewski, ZR and Baugh, J},
  journal={Applied Physics Letters},
  volume={122},
  number={1},
  year={2023},
  publisher={AIP Publishing}
}

@article{zhang2020interface,
  title={Interface Engineering and Epitaxial Growth of Single-Crystalline Aluminum Films on Semiconductors},
  author={Zhang, Kedong and Xia, Shunji and Li, Chen and Pan, Jiahui and Ding, Yuanfeng and Lu, Ming-Hui and Lu, Hong and Chen, Yan-Feng},
  journal={Advanced Materials Interfaces},
  volume={7},
  number={16},
  pages={2000572},
  year={2020},
  publisher={Wiley Online Library}
}

@phdthesis{johnson1995optical,
  title={Optical bandgap thermometry in molecular beam epitaxy},
  author={Johnson, Shane R},
  year={1995},
  school={University of British Columbia}
}

@article{averbeck1991oxide,
  title={Oxide desorption from InP under stabilizing pressures of P2 or As4},
  author={Averbeck, R and Riechert, H and Schl{\"o}tterer, H and Weimann, G},
  journal={Applied physics letters},
  volume={59},
  number={14},
  pages={1732--1734},
  year={1991},
  publisher={American Institute of Physics}
}

@article{dmitriev2021transformation,
  title={Transformation of the InP (001) surface upon annealing in an arsenic flux},
  author={Dmitriev, Dmitriy V and Kolosovsky, Danil A and Gavrilova, Tatyana A and Gutakovskii, Anton K and Toropov, Alexander I and Zhuravlev, Konstantin S},
  journal={Surface Science},
  volume={710},
  pages={121861},
  year={2021},
  publisher={Elsevier}
}

@article{tournet2016growth,
  title={Growth and characterization of epitaxial aluminum layers on gallium-arsenide substrates for superconducting quantum bits},
  author={Tournet, Julie and Gosselink, Denise and Miao, Guo-Xing and Jaikissoon, Marc and Langenberg, Deler and McConkey, Thomas G and Mariantoni, Matteo and Wasilewski, Zbigniew R},
  journal={Superconductor Science and Technology},
  volume={29},
  number={6},
  pages={064004},
  year={2016},
  publisher={IOP Publishing}
}

@article{leroy2016control,
  title={How to control solid state dewetting: A short review},
  author={Leroy, Fr{\'e}d{\'e}ric and Cheynis, F and Almadori, Y and Curiotto, S and Trautmann, M and Barb{\'e}, JC and M{\"u}ller, P and others},
  journal={Surface Science Reports},
  volume={71},
  number={2},
  pages={391--409},
  year={2016},
  publisher={Elsevier}
}

@article{wang2015sharp,
  title={Sharp interface model for solid-state dewetting problems with weakly anisotropic surface energies},
  author={Wang, Yan and Jiang, Wei and Bao, Weizhu and Srolovitz, David J},
  journal={Physical Review B},
  volume={91},
  number={4},
  pages={045303},
  year={2015},
  publisher={APS}
}

@article{zucker2013model,
  title={A model for solid-state dewetting of a fully-faceted thin film},
  author={Zucker, Rachel V and Kim, Gye Hyun and Carter, W Craig and Thompson, Carl V},
  journal={Comptes Rendus. Physique},
  volume={14},
  number={7},
  pages={564--577},
  year={2013}
}

@article{hieke2017annealing,
  title={Annealing induced void formation in epitaxial Al thin films on sapphire ($\alpha$-Al2O3)},
  author={Hieke, Stefan Werner and Dehm, Gerhard and Scheu, Christina},
  journal={Acta Materialia},
  volume={140},
  pages={355--365},
  year={2017},
  publisher={Elsevier}
}

@article{hieke2017microstructural,
  title={Microstructural evolution and solid state dewetting of epitaxial Al thin films on sapphire ($\alpha$-Al2O3)},
  author={Hieke, Stefan Werner and Breitbach, Benjamin and Dehm, Gerhard and Scheu, Christina},
  journal={Acta Materialia},
  volume={133},
  pages={356--366},
  year={2017},
  publisher={Elsevier}
}

@article{klinger2017more,
  title={More features, more tocrols, more CrysTBox},
  author={Klinger, Miloslav},
  journal={Applied Crystallography},
  volume={50},
  number={4},
  pages={1226--1234},
  year={2017},
  publisher={International Union of Crystallography}
}

@article{yasaka2010x,
  title={X-ray thin-film measurement techniques},
  author={Yasaka, Miho and others},
  journal={The Rigaku Journal},
  volume={26},
  number={2},
  pages={1--9},
  year={2010}
}

@article{poulopoulos2015growth,
  title={Growth and optical properties of nanocrystalline titania films for Optoelectronics and photovoltaics},
  author={Poulopoulos, P and Grammatikopoulos, S and Trachylis, D and Bissas, G and Dragatsikas, I and Velgakis, MJ and Politis, C},
  journal={Journal of Surfaces and Interfaces of Materials},
  volume={3},
  number={1},
  pages={52--59},
  year={2015},
  publisher={American Scientific Publishers}
}

@article{aid1999atomic,
  title={Atomic structure of the (2$\times$ 4) In0. 53Ga0. 47As/InP (001) reconstructed surface. A study of average strain and growth temperature effects on the indium segregation},
  author={A{\i}d, K and Garreau, Y and Sauvage-Simkin, M and Pinchaux, R},
  journal={Surface Science},
  volume={425},
  number={2-3},
  pages={165--173},
  year={1999},
  publisher={Elsevier}
}

@article{ozanyan1997situ,
  title={In situ monitoring of the surface reconstructions on InP (001) prepared by molecular beam epitaxy},
  author={Ozanyan, KB and Parbrook, PJ and Hopkinson, M and Whitehouse, CR and Sobiesierski, Zbigniew and Westwood, David I},
  journal={Journal of applied physics},
  volume={82},
  number={1},
  pages={474--476},
  year={1997},
  publisher={American Institute of Physics}
}

@article{labella2000reflection,
  title={Reflection high-energy electron diffraction and scanning tunneling microscopy study of InP (001) surface reconstructions},
  author={LaBella, VP and Ding, Z and Bullock, DW and Emery, C and Thibado, PM},
  journal={Journal of Vacuum Science \& Technology A: Vacuum, Surfaces, and Films},
  volume={18},
  number={4},
  pages={1492--1496},
  year={2000},
  publisher={American Vacuum Society}
}

@article{millunchick2004surface,
  title={Surface reconstructions of InGaAs alloys},
  author={Millunchick, J Mirecki and Riposan, A and Dall, BJ and Pearson, Chris and Orr, BG},
  journal={Surface science},
  volume={550},
  number={1-3},
  pages={1--7},
  year={2004},
  publisher={Elsevier}
}

\end{document}